**A Remarkably Accurate Predictor of Sunspot Cycle Amplitude.**

Peter Foukal, Nahant, MA 01908 (pvfoukal@comcast.net)

**Abstract**

The slopes of the linear relations between sunspot and white light (WL) facular areas at the onset of sunspot Cycles 12-21 correlate well with the amplitudes of those cycles between 1878-1980 (Brown and Evans, 1980). We use continuum images from the SOHO Michelson Doppler Imager and SDO Heliospheric Magnetic Imager to show that the relation holds also for Cycles 24 and 25. The amplitudes of Cycles 12-21 and 24 calculated using this relation agree with the observed amplitudes to within +/- 4% rms. It also enabled us in 2022 to correctly predict a larger Cycle 25 than estimated by the International Prediction Panel, 3 years before maximum. The technique offers an objective, physically based predictor of cycle amplitudes 3-4 years ahead of their maxima, given a stable source of continuum full disk photospheric images.

## 1. Introduction

Predictions of the amplitude of the next sunspot cycle are one of the most important practical applications of solar research and astronomy. The dynamics of the magnetic cycle is still too uncertain to provide a physical basis for reliable predictions so estimates of the cycle amplitude and timing are derived largely from empirical or semi-empirical models (e.g Schatten, 2005; Svalgaard, Cliver and Kamide, 2005; McIntosh et al, 2020; see e.g. Pesnell, 2020; Petrovay, 2020 for recent reviews).

This study builds on the finding (Brown and Evans, 1980) that the slopes, $dA_s/dA_f$, of the linear relation between sunspot and WL facular areas early in a cycle are closely related to the cycle's amplitude. This relationship was originally discovered in analysis of the measurements of spots and WL faculae made by the Royal Greenwich Observatory (RGO) between 1874 and 1976. Brown and Evans used it to predict the amplitude of Cycle 21 to within 6% of its observed amplitude.

The RGO program (Royal Observatory, 1973) ended in 1976 and this relationship was not exploited to predict the amplitudes of subsequent cycles. Emulsion inhomogeneity and plate exposure variations make measurement of WL facular areas difficult on digitizations of the Mt Wilson Observatory continuum spectro - heliograms and Kodaikanal Observatory photo - heliograms, the two most complete sets of full disk continuum images available after 1976.

More recently, excellent daily full disk continuum images were obtained first from the Michelson Doppler Imager (MDI) on the Solar and Heliospheric Observatory (SOHO) beginning in 2001 (Scherrer et al., 1995) and then from the Heliospheric and Magnetic Imager (HMI) on the Solar Dynamics Observatory (SDO) since 2011 (Scherrer et al., 2012). These enable us to show that the relationship holds also for Cycles 24 and 25.

## 2. Data and analysis

The MDI images were obtained in narrow - band (94 mÅ) quasi-continuum in the wing of the Ni 6768 Å line. The 75 mÅ HMI continuum passband was located in the wing of the Fe I 6173 Å line. Both data sets had been corrected for limb darkening by the P.I. teams. These passbands showed faculae only nearer the limb than a heliocentric angle of approximately 50 degrees, in agreement with their appearance on the continuum images used in the RGO measurements. The resolution, uniformity, and dynamic range of the space-borne imagery (which shows the photospheric granulation) are superior to the definition of the photographic plates used by RGO. A representative HMI image is shown in Figure 1.

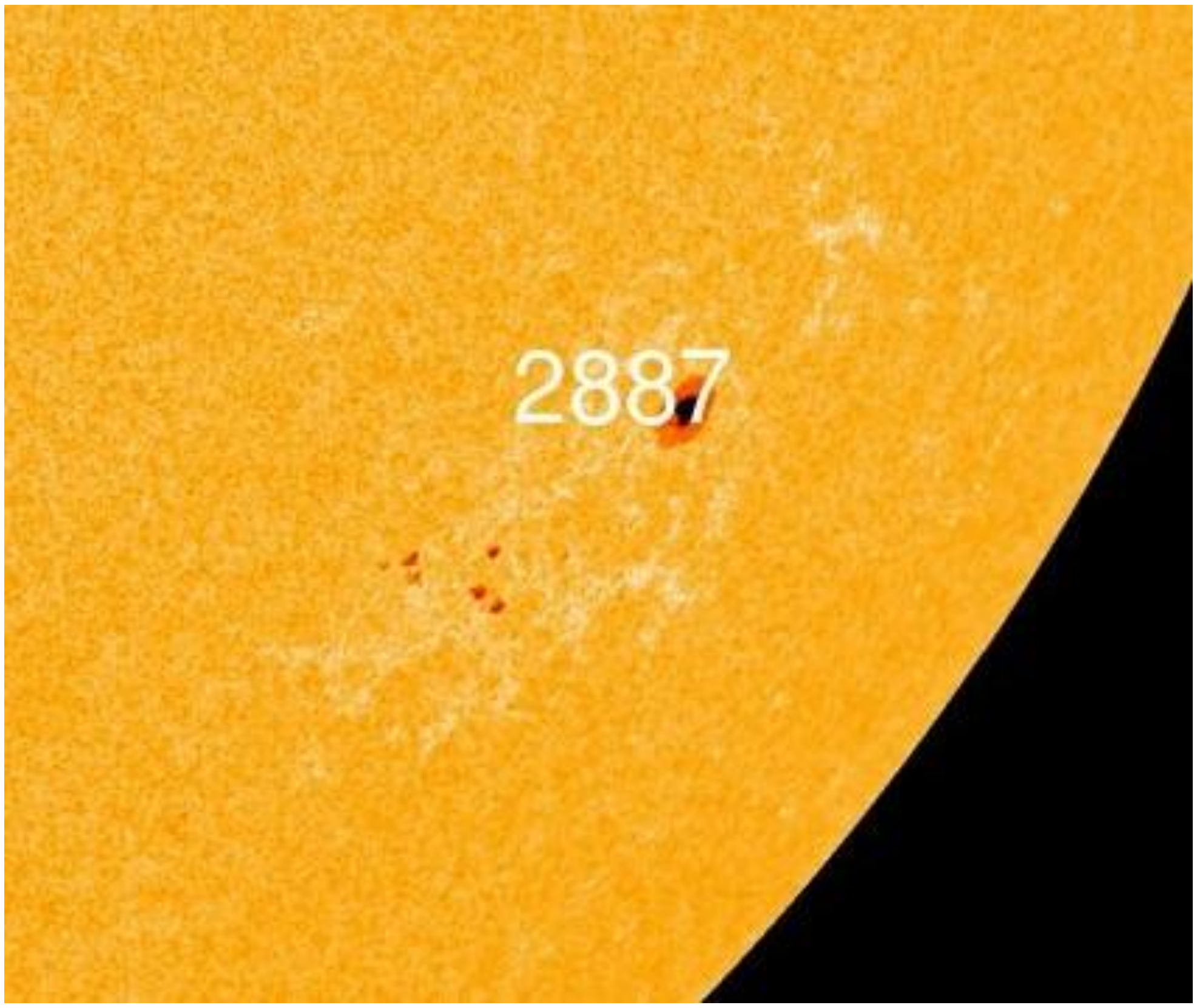

Figure 1 Close-up of limb faculae imaged by HMI on 1 November 2011.

Four equally-spaced MDI images per month for each of the 24 months during the two first years 2009-2010 of Cycle 24 were downloaded and printed. The same procedure was followed for the two first years 2020-2021 of Cycle 25. This sampling is adequate given the high quality of the MDI and HMI images, the lifetimes of large WL faculae and spots that determine most of these features' total area, and the 13-month smoothing of our data.

The projected areas of spots and faculae on the 130 mm (MDI) and 190 mm (HMI) diameter printed images were measured by counting squares using a magnifier with a reticle grid of 0.5 mm spacing, and corrected for foreshortening. This is similar to the procedure used by the RGO between 1874-1976. Monthly means of these corrected spot and facular areas were then smoothed with an evenly weighted 13-month running mean, following the method used by Brown and Evans (1980).

3. **Results and errors**

A plot of these running means of spot ($A_s$) versus facular ($A_f$) areas for Cycle 24 is shown in Figure 2. We see that the spot and facular areas are linearly related ($R^2$ = 0.88) with a slope of 0.24 in this cycle. A similar plot is shown in Figure 3 for Cycle 25. Again, the relation is linear ($R^2$ = 0.97) but the slope of 0.61 is greater. Brown and Evans (1980) showed that such linear relationships were characteristic of approximately the first two years of all the cycles measured by RGO.

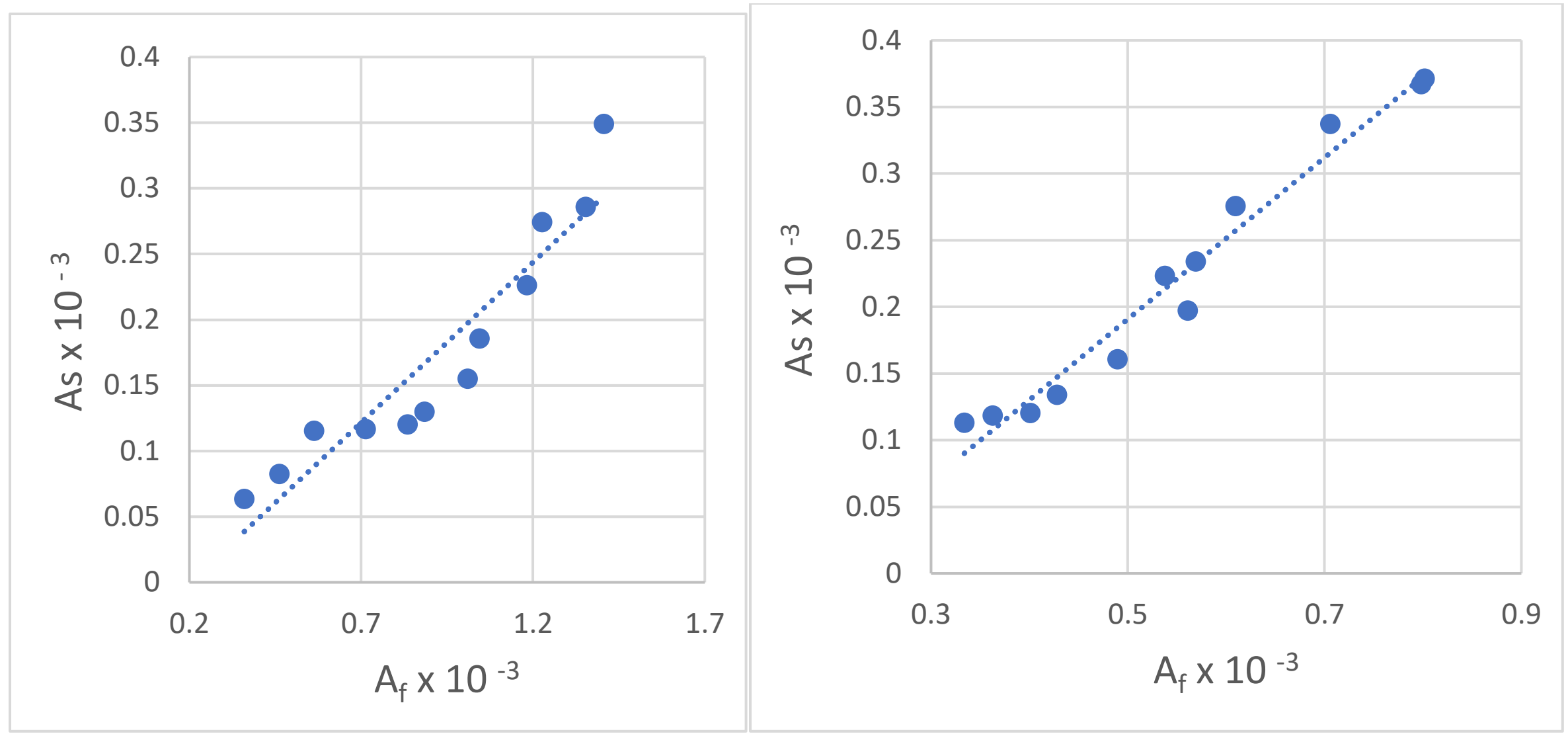

Figure 2 $A_s$ versus $A_f$ (μh) for 2009 – 2010. Figure 3 $A_s$ versus $A_f$ (μh) for 2020 – 2021.

Figure 4 shows the linear relationship ($R^2$ = 0.95) between the slopes of the Cycles 12-21 derived by Brown and Evans, and the peak 13 - month smoothed sunspot number, $R_m$-max, of those cycles. The slope of 0.24 for Cycle 24 is indicated by an inverted triangle at the bottom of the plot. It corresponds to a peak cycle amplitude of approximately 115, which agrees with its 13- month smoothed amplitude of 115 achieved in 2014. The values of $R_m$–max used in this study are on the re-calibrated International Sunspot Number scale (Clette et al., 2016). They are about 35% higher than those on the earlier Zurich sunspot number scale used by Brown and Evans (1980).

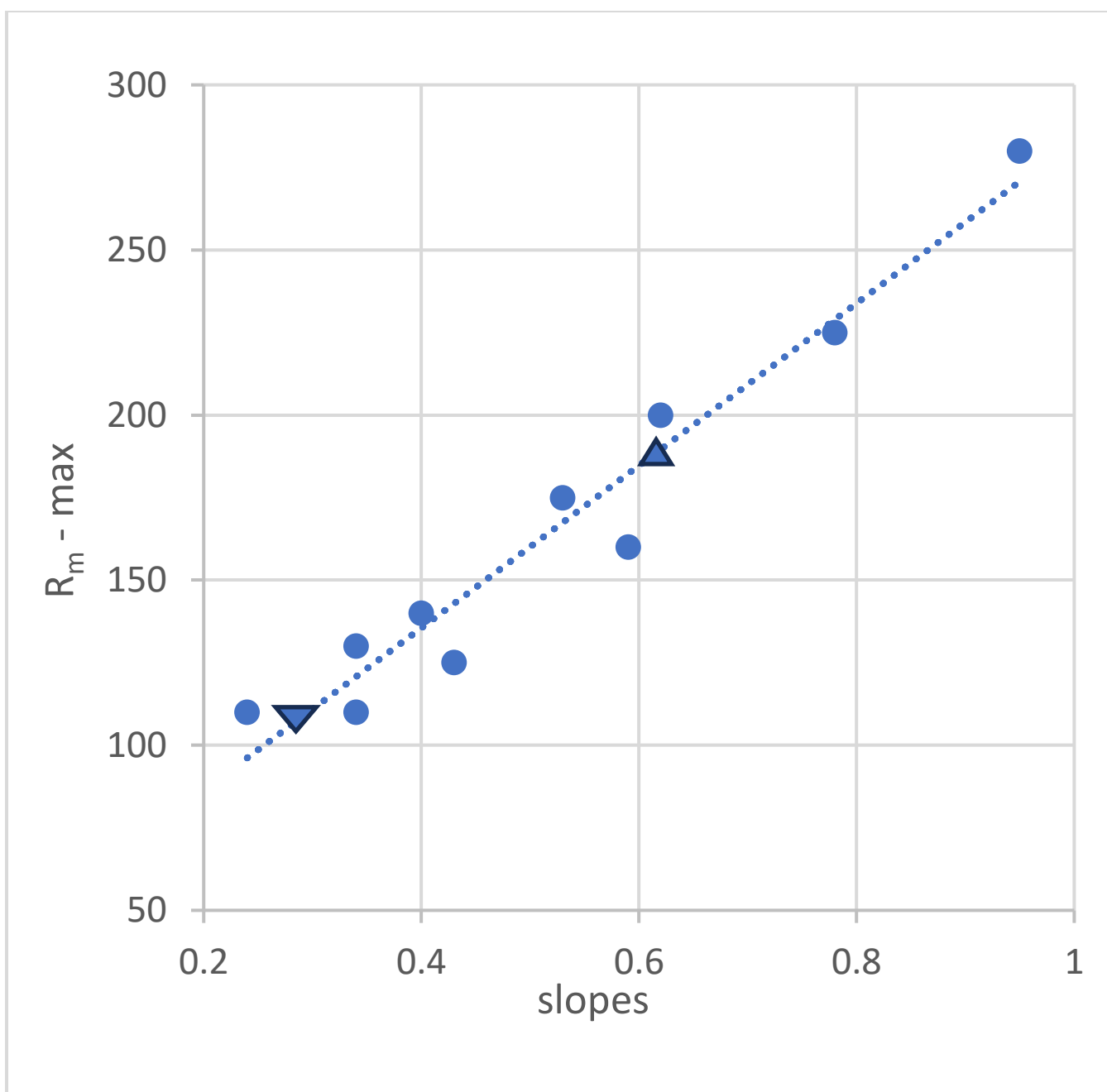

Figure 4 $R_m$-max versus slopes $dA_s/dA_f$. The dotted line represents a linear fit to the 10 points derived by Brown and Evans (1980), shown as filled circles. The slope values derived here for Cycles 24 and 25 are shown as the inverted and upright triangles, respectively.

The slope of 0.61 measured here from HMI data for Cycle 25 yields a peak amplitude of $R_m$–max = 185, about 16% higher than the actual $R_m$-max achieved in mid-2025.

Györi (2012) reports that the MDI spot and facular areas exceed the HMI values because of the lower angular resolution of the MDI images. His relations for HMI corrected areas relative to MDI values are: $A_s$ MDI = ($A_s$ HMI - 4)/0.91, and $A_f$ MDI = ($A_f$ MDI – 260)/0.67. Using these we obtain a slope of 0.44 in Figure 3 and a lower $R_m$ -max ~ 145.

That analysis compared MDI data obtained during the decline of Cycle 23 with different structures observed by HMI during the rise of Cycle 24. We compared instead the areas obtained when the *same* structures are observed co-temporally during the August – December 2010 overlap period between the MDI and HMI missions. The results shown in Figures 5 and 6 yield the transformation relations: As MDI = (As HMI - 2)/0.88 ($r^2$ = 0.93) and $A_f$ MDI = ($A_f$ HMI – 4)/0.87 ($r^2$ = 0.62). The comparison of projected rather than corrected areas used in these Figures avoids error from unnecessary foreshortening corrections.

Our transformations essentially cancel and make no significant difference to the slope of Figure 3. This is consistent with the blurring interpretation of the larger MDI areas since the RGO method of area measurement we adopted for consistency introduces similar blurring for the MDI and HMI images. It is also consistent with the report by Jha et al. (2019) of good agreement between MDI spot areas and those measured from Kodaikanal Solar Observatory photo - heliograms when both are degraded to similar angular resolution.

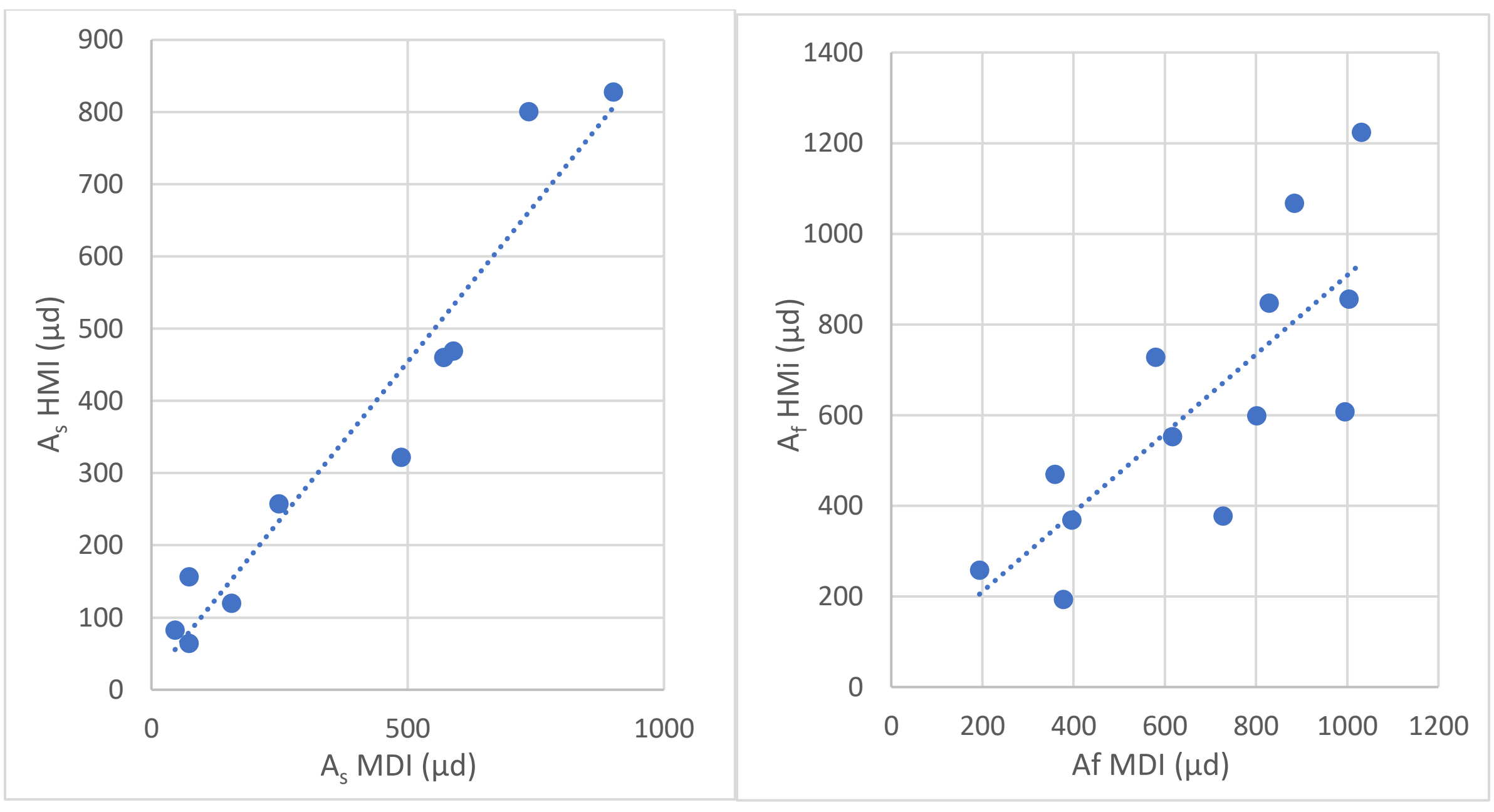

Figure 5 $A_s$ HMI versus MDI (millionths of disk). Figure 6 $A_f$ HMI versus MDI (millionths of disk).

The corrections reported by Györi (2012) apply to comparison of the MDI images with those obtained from HMI at significantly higher angular area resolution so they cannot be applied directly to our measurements. Unfortunately, the low activity level during the short overlap period precludes defining our facular correction accurately enough to decide confidently whether it can account for the 16% difference between calculated and actual $R_m$ – max in Cycle 25.

## 4. Discussion and Conclusions

We have shown that the linear relations between spot and facular areas, $A_s$ and $A_f$, and between the slopes $dA_s/dA_f$ and cycle amplitudes $R_m$ – max, first reported for the onsets of Cycles 12-21 by Brown and Evans (1980) hold also for Cycles 24 and 25. The calculated and actual amplitudes of Cycles 12-24 (see Table I) agree to within +/- 4 % rms. This represents unprecedented accuracy over an unequaled span of 1 ½ centuries.

Table I

Comparison of calculated and observed 13-month smoothed amplitudes of Cycles 12-21 and 24, 25.

| Cycle | Calc | Obs | Calc – Obs |
|---|---|---|---|
| 12 | 105 | 100 | 5 |
| 13 | 125 | 125 | 0 |
| 14 | 85 | 85 | 0 |
| 15 | 150 | 140 | 10 |
| 16 | 105 | 105 | 0 |
| 17 | 170 | 160 | 10 |
| 18 | 210 | 205 | 5 |
| 19 | 285 | 285 | 0 |
| 20 | 155 | 155 | 0 |
| 21 | 210 | 225 | -15 |
| 24 | 115 | 120 | -5 |
| 25 | 185 | 160 | 25 |

The 16% over-estimate in our 2022 prediction of the Cycle 25 amplitude of $R_m$ – max = 185 (Foukal, 2023) was probably caused mainly by the uncertain correction between facular areas measured from the high - resolution HMI images, as compared to those we measured from the MDI and those measured by RGO from ground-based photo - heliograms of similar resolution. Despite this uncertainty we were able to correctly predict a Cycle 25 amplitude larger than the prediction by the International Prediction Panel (IPP) at that epoch (e.g. Miesch, 2023)

The approximately 3 ½ year advance notice of our Cycle 25 prediction is shorter than provided by some of the 50 or more other methods used by the IPP (see e.g. Miesch, 2023 for an overview of Cycle 25 predictions). Nor does this method predict the timing of maxima, although a useful relation between $R_m$ - max and the cycle rise time (Waldmeier, 1978) can be helpful in this respect.

The method described here is empirical but its physical basis can be understood in terms of the evidence on the relative areas of spots and faculae in the Sun and other late-type stars (e.g. Foukal, 1994; 1998). Dark spots are produced by magnetic flux tubes whose areas lie at the upper end of the photospheric flux tube size spectrum. The WL faculae lie at the lowest end (e.g. Spruit, 1976).

The increase in the slope $dA_s/dA_f$ between Cycles 24 and 25 seen in Figures 2 and 3 therefore corresponds to a shift in the size spectrum of photospheric magnetic fields towards more power at the larger scales, as the cycle amplitude increases. Such a tendency to shift dynamo power towards lower spatial wavenumbers as magnetic flux output increases has been recognized in observations of large spots on younger late-type stars (e.g. Radick, Lockwood and Baliunas, 1990) and in dynamo calculations (e.g. Cattaneo, Cieunh and Hughes, 1990).

The relationship studied here between $A_s$ and $A_f$ does not hold as well between sunspots and bright chromospheric plages (Foukal, 1998) because these cover not only the most slender flux tubes observed as WL faculae, but also pores and small spots. This decreases the sensitivity of the present diagnostic. So daily Ca K plage areas available for Cycles 22 and 23 (e.g. Foukal et al., 2009) cannot be used to substitute the missing facular data for those cycles. But extension of the method to Cycle 23 using the enhanced contrast continuum images from the Kanzelhöhe Observatory (Pötzi, 2025) deserves investigation.

**Acknowledgements**

This study was facilitated by availability of the MDI and HMI images obtained by the Stanford University and Lockheed Corporation Solar Groups on Spaceweather.com, and on the Debrecen Observatory site (fenyi.solarobs.hunren.hu). I am grateful to Frédéric Clette and Mark Miesch for information on the SC 25 Prediction Panel, to the late Leif Svalgaard for drawing my attention to Waldmeier's rule and to Neil Sheeley and an anonymous referee for comments on an original version of this paper. I thank Tom Williams for drawing my attention to the Spaceweather.com site and Jack Martin for facilitating access to the RGO archive at Cambridge University; I am grateful to the archivist, Dr. Emma Saunders, for her kind assistance.

The author declares that he has no competing interests.